\documentclass[aps,pra,reprint,showpacs,longbibliography]{revtex4-1}
\usepackage[latin9]{inputenc}
\usepackage{prettyref}
\usepackage{amsmath}
\usepackage{amssymb}
\usepackage{graphicx}

\makeatletter
\usepackage{graphicx}

\newcommand*{\pr}[1]{\mathcal{#1}}
\newcommand*{\refneq}[1]{(\ref{#1})}
\newcommand*{\refeq}[1]{Eq.\ (\ref{#1})}
\newrefformat{eq}{Eq.\ (\ref{#1})}

\makeatother

\begin{document}
\global\long\def\pr#1{\mathcal{#1}}

\title{Long-lived states with well-defined spins in spin-$1/2$ homogeneous
Bose gases}

\author{Vladimir A. Yurovsky}

\affiliation{School of Chemistry, Tel Aviv University, 6997801 Tel Aviv, Israel}

\date{\today}
\begin{abstract}
Many-body eigenfunctions of the total spin operator can be constructed
from the spin and spatial wavefunctions with non-trivial permutation
symmetries. Spin-dependent interactions can lead to relaxation of
the spin eigenstates to the thermal equilibrium. A mechanism that
stabilizes the many-body entangled states is proposed here. Surprisingly,
in spite coupling with the chaotic motion of the spatial degrees of
freedom, the spin relaxation can be suppressed by destructive quantum
interference due to spherical vector and tensor terms of the spin-dependent
interactions. Tuning the scattering lengths by the method of Feshbach
resonances, readily available in cold atomic labs, can enhance the
relaxation timescales by several orders of magnitude.
\end{abstract}


\maketitle

\section*{Introduction}

A non-degenerate gas of interacting particles far from critical points
is generally regarded as one of the most pronounced representatives
of chaotic systems. According to the eigenstate thermalization hypothesis
\cite{deutsch1991,srednicki1994}, expectation values of observables
in the gas eigenstates coincide with microcanonical expectation values.
The expectation values relax to the equilibrium after about 3 collisions,
as demonstrated by numerical simulations and experiments \cite{monroe1993,wu1996,kinoshita2006}. 

Gases of spinor particles are attracting increasing attention starting
from the first experimental \cite{myatt1997,stamper1998} and theoretical
\cite{ho1998,ohmi1998} works (see book \cite{pitaevskii}, reviews
\cite{stamper2013,guan2013} and references therein). Such gases can
be described in two ways. In the first, conventional, description,
each particle acquires an additional degree of freedom --- the spin
projection $s_{z}$, which can have values $\pm\frac{1}{2}$, for
spin-$\frac{1}{2}$ particles. It can be either the projection of
a real, physical, angular momentum, or it can be attributed to internal
states of particles (e.g. hyperfine states of atoms). In the last
case, the particles can be either bosons or fermions, with no relation
to their spins. The sum of the particle spin projections, the total
spin projection, is conserved in the absence of spin-changing collisions,
being related to occupations of the spin states. Then the gas is a
mixture of the gases of particles in the given spin state, which relax
to the thermal equilibrium with the same temperature.

Another description of spinor gases is based on collective spin and
spatial wavefunctions. It is a generalization of the well-known representation
of a two-electron wavefunction as a product of permutation-symmetric
spatial and antisymmetric spin wavefunctions for the singlet state
or antisymmetric spatial and symmetric spin ones for the triplet state.
The singlet and triplet states have different energies due to the
coulombic interaction between electrons.

The symmetric and antisymmetric functions are examples of irreducible
representations of the symmetric group \cite{hamermesh,kaplan,pauncz_symmetric}.
Spatial and spin wavefunctions of $N$-body systems with $N>2$ can
belong to multidimensional, non-Abelian, irreducible representations,
when permutations transform a function to a superposition of the representation
functions. In the case of spin- $\frac{1}{2}$ particles, the representations
are associated with the total spin $S$. The total wavefunctions with
the correct bosonic or fermionic permutation symmetry are expressed
as a sum of products of the spin and spatial functions \cite{kaplan,pauncz_symmetric}.
The only one-dimensional representations, the symmetric and antisymmetric
functions, are associated with $S=N/2$ and $S=0$. Spin-independent
interactions between particles split energies of states with different
$S$, as shown by Heitler \cite{heitler1927}. The states with well-defined
total spin are used in quantum chemistry (see \cite{kaplan,pauncz_symmetric})
and were applied to spinor gases \cite{lieb1962,yang1967,sutherland1968,guan2009,yang2009,gorshkov2010,fang2011,daily2012,harshman2014,yurovsky2014,yurovsky2015,yurovsky2015a,harshman2016,harshman2016a}.
Many-body entanglement of such states can by employed for quantum
computing \cite{jordan2010}. Another example of states with defined
spins is the Dicke collective state of of two-level atoms or molecules
coupled by a single mode of the electromagnetic field \cite{dicke1954}
(see also the recent work \cite{sela2014} and the references therein). 

In the case of spin-independent interactions, the total spin is conserved,
the gas can be created in a state with given $S$, and will not relax
to the thermal equilibrium, which corresponds to a mixture of states
with different $S$. The present work analyses the relaxation of such
states due to spin-dependent interactions between particles. It demonstrates
that, in spite of coupling to chaotic spatial motion, the spin-relaxation
can be suppressed due to quantum interference tuned by a Feshbach
resonance. The relaxation time-scales can be also enhanced in non-equilibrium
ways \cite{das2010,roy2015,cemkeser2015} using time-dependent perturbations.
Relaxation of the Dicke states can be suppressed due to interaction
with cavity modes \cite{sela2016}. 

The paper has the following structure. \prettyref{sec:HamWFperm}
describes spin-dependent interactions and permutation-symmetric wavefunctions.
The Berry's conjecture \cite{berry1977} and eigenstate thermalization
\cite{srednicki1994} methods for description of the chaos in the
spatial degrees of freedom are generalized in Sec. \ref{sec:QuantChaos}
to the states with well-defined spins. In Sec. \ref{sec:Decay} these
methods are used for calculation of non-diagonal matrix elements and
relaxation rates, in a combination with symmetric group methods and
the sum rules \cite{yurovsky2015,yurovsky2015a}.

\section{The Hamiltonian, wavefunctions, and permutation symmetry\label{sec:HamWFperm}}

A general interaction, which does not change the spin projection,
is a sum of interactions of particles in each combination of the two
spin states, $\uparrow$ and $\downarrow$,
\begin{equation}
\hat{V}_{\mathrm{spin}}=\frac{g_{Dd}^{\upuparrows}}{2}\hat{V}_{\upuparrows}+\frac{g_{Dd}^{\downdownarrows}}{2}\hat{V}_{\downdownarrows}+g_{Dd}^{\uparrow\downarrow}\hat{V}_{\uparrow\downarrow}.\label{eq:Vspin}
\end{equation}
 Whenever the thermal wavelength for the temperature $T$ 
\begin{equation}
\lambda_{T}=[2\pi\hbar^{2}/(mT)]^{1/2}\label{eq:lambdaT}
\end{equation}
 substantially exceeds the interaction range, the interactions can
be approximated by the zero-range ones 
\[
\hat{V}_{\upuparrows}=\sum_{j\neq j'}\delta(\mathbf{r}_{j}-\mathbf{r}_{j'})|\uparrow(j)\rangle|\uparrow(j')\rangle\langle\uparrow(j)|\langle\uparrow(j')|
\]
\[
\hat{V}_{\downdownarrows}=\sum_{j\neq j'}\delta(\mathbf{r}_{j}-\mathbf{r}_{j'})|\downarrow(j)\rangle|\downarrow(j')\rangle\langle\downarrow(j)|\langle\downarrow(j')|
\]
\[
\hat{V}_{\uparrow\downarrow}=\sum_{j\neq j'}\delta(\mathbf{r}_{j}-\mathbf{r}_{j'})|\uparrow(j)\rangle|\downarrow(j')\rangle\langle\uparrow(j)|\langle\downarrow(j')|
\]
{[}note the double-counting the particle pairs in $\hat{V}_{\upuparrows}$
and $\hat{V}_{\downdownarrows}$, which is compensated by the factors
$\frac{1}{2}$ in \prettyref{eq:Vspin}{]}. The particle coordinates
$\mathbf{r}_{j}$ are vectors in $D$-dimensional space ($D=2$ or
3). In the two-dimensional (2D) case, the motion in the third (axial)
dimension is confined by a harmonic potential with the frequency $\omega_{\mathrm{conf}}$
and the two-dimensional gas can be formed at sufficiently low temperature
$T<\hbar\omega_{\mathrm{conf}}$. In certain situations, the two-
and three-dimensional $\delta$-functions should be renormalized in
order to eliminate divergences. The interaction strengths
\begin{equation}
g_{3d}^{\sigma\sigma'}=4\pi\hbar^{2}\frac{a_{\sigma\sigma'}}{m},\quad g_{2d}^{\sigma\sigma'}=\sqrt{\frac{m\omega_{\mathrm{conf}}}{2\pi\hbar}}g_{3d}^{\sigma\sigma'},\label{eq:ga3d2d}
\end{equation}
where $m$ is the boson's mass, are proportional to the elastic scattering
lengths $a_{\sigma\sigma'}$.

The interactions $\hat{V}_{\upuparrows}$, $\hat{V}_{\downdownarrows}$,
$\hat{V}_{\uparrow\downarrow}$, and, therefore, $\hat{V}_{\mathrm{spin}}$
in \prettyref{eq:Vspin} can be expanded in terms of irreducible spherical
tensors \cite{yurovsky2015a}
\begin{multline}
\hat{V}_{\mathrm{spin}}=\frac{1}{2}\Bigl[g_{Dd}\hat{V}+(g_{Dd}^{\upuparrows}-g_{Dd}^{\downdownarrows})\hat{V}_{0}\\
+\sqrt{\frac{2}{3}}(g_{Dd}^{\upuparrows}+g_{Dd}^{\downdownarrows}-2g_{Dd}^{\uparrow\downarrow})\hat{V}_{0}^{(2)}\Bigr].\label{eq:VspinSph}
\end{multline}
The spherical scalar interaction 
\[
\frac{1}{2}g_{Dd}\hat{V}=\frac{1}{2}g_{Dd}\sum_{j\neq j'}\delta(\mathbf{r}_{j}-\mathbf{r}_{j'})
\]
with the interaction strength $g_{Dd}=(g_{Dd}^{\upuparrows}+g_{Dd}^{\downdownarrows}+g_{Dd}^{\uparrow\downarrow})/3$
provides the spin-independent interaction between particles. If all
scattering lengths have the same value, the spin-dependent parts of
the interaction vanish and the Hamiltonian of $N$ indistinguishable
spin-$\frac{1}{2}$ bosons has the form
\[
\hat{H}=\hat{H}_{0}+\frac{1}{2}g_{Dd}\hat{V}
\]
where
\[
\hat{H}_{0}=\frac{1}{2m}\sum_{j}\mathbf{\hat{p}}_{j}^{2}
\]
is the kinetic energy and $\mathbf{\hat{p}}_{j}$ are the momentum
operators.

Since $\hat{H}$ is invariant over independent permutations of the
particle spins and coordinates and commutes with the operators of
the total spin $\hat{S}$ and its projection $\hat{S}_{z}$, the eigenfunctions
can be expressed as \cite{kaplan,pauncz_symmetric,yurovsky2015} 
\begin{equation}
\Psi_{nS_{z}}^{(S)}=f_{S}^{-1/2}\sum_{t}\Phi_{tn}^{(S)}\Xi_{tS_{z}}^{(S)},\label{eq:PsiPhiXi}
\end{equation}
where the spatial $\Phi_{tn}^{(S)}$ and spin $\Xi_{tS_{z}}^{(S)}$
wavefunctions form bases of irreducible representations of the symmetric
group $\pr S_{N}$ of permutations of $N$ symbols \cite{hamermesh,kaplan,pauncz_symmetric}.
The representations are associated with the two-row Young diagrams
$\lambda=[N/2+S,N/2-S]$ and have dimensions
\[
f_{S}=\frac{N!(2S+1)}{(N/2+S+1)!(N/2-S)!}.
\]
The basis functions within the representations are labeled by the
standard Young tableaux $t$ of the shape $\lambda$. A permutation
$\pr P$ of the particles transforms each function to a linear combination
of functions in the same representation,
\begin{equation}
\pr P\Phi_{tn}^{(S)}=\sum_{t'}D_{t't}^{[\lambda]}(\pr P)\Phi_{t'n}^{(S)},\quad\pr P\Xi_{tS_{z}}^{(S)}=\sum_{t'}D_{t't}^{[\lambda]}(\pr P)\Xi_{t'S_{z}}^{(S)}.\label{eq:PermPhiXi}
\end{equation}
Here $D_{t't}^{[\lambda]}(\pr P)$ are the Young orthogonal matrices.
Their properties \cite{hamermesh,kaplan,pauncz_symmetric} provide
the correct bosonic transformation $\pr P\Psi_{nl}^{(S)}=\Psi_{nl}^{(S)}$
for the total wavefunction \eqref{eq:PsiPhiXi}. 

The explicit form of the spin wavefunctions \cite{yurovsky2013} is
not used here. Their orthonormality 
\[
\left\langle \Xi_{t'S'_{z}}^{(S')}|\Xi_{tS_{z}}^{(S)}\right\rangle =\delta_{S'S}\delta_{t't}\delta_{S'_{z}S_{z}}
\]
leads to the Schr\"odinger equation for the spatial wavefunctions
\begin{equation}
\hat{H}\Phi_{tn}^{(S)}=E_{n}^{(S)}\Phi_{tn}^{(S)}\label{eq:SchrPhi}
\end{equation}
(all wavefunctions within an irreducible representation are energy-degenerate,
according to the Wigner theorem).

\section{Quantum-chaotic wavefunctions with defined total spins\label{sec:QuantChaos}}

Consider $N$ spin-$\frac{1}{2}$ bosons in a periodic box with incommensurable
dimensions. The box can be either three-dimensional (3D) of the volume
$L^{3}$ or 2D of the square $L^{2}$ with tight confinement axial
dimension. A tight confinement in two directions can lead to a homogeneous
one-dimensional gas (see \cite{yurovsky2008b}), which is an integrable
system and is not considered here. In contrast, homogeneous 2D and
3D gases can demonstrate chaotic behavior at the sufficiently high
energy-density of states, as assumed in the first consideration of
eigenstate thermalization by Deutsch \cite{deutsch1991}. According
to the quantitative criteria \cite{altshuler1997,jacquod1997,flambaum1997},
based on analyses of delocalization in the Fock space, many-body systems
become chaotic when the interaction matrix elements exceed the energy
spacing of directly-coupled many-body states. In the 3D case, the
matrix elements are $g_{3d}/L^{3}$. The two-body interactions couple
states with different momenta of any of $N(N-1)/2$ pairs of particles
and conserved center-of-mass momentum. Then the energy density of
coupled states will be $N(N-1)/2$ times the energy density of relative-motion
states $L^{3}m^{3/2}T^{1/2}/(4\pi^{2}\hbar^{3})$. This leads to the
criterion of chaos 
\begin{equation}
a_{S}\equiv\frac{a_{\upuparrows}+a_{\downdownarrows}+a_{\uparrow\downarrow}}{3}>\frac{\sqrt{2\pi}\lambda_{T}}{N^{2}},\label{eq:aSchaos3d}
\end{equation}
where the thermal wavelength $\lambda_{T}$ is given by \prettyref{eq:lambdaT}.
This criterion do not contradict to the sufficient condition \cite{srednicki1994}
for the hard-sphere gas, based on calculations for chaotic billiards,
that chaos appears when the particle radius exceeds the thermal wavelength.
Compared to this condition, the present criterion reduces the threshold
temperature by the factor $N^{4}$. In the 2D case, the energy density
of relative-motion states is $mL^{2}/(4\pi\hbar^{2})$ and the matrix
elements are $g_{2d}/L^{2}$. Then the condition of chaos will be
\begin{equation}
a_{S}>2\sqrt{2\pi\hbar/(m\omega_{\mathrm{conf}})}/N^{2},\label{eq:chaos2d}
\end{equation}
where the square root is, up to a factor, the confinement range. 

Under the conditions of chaos, the spatial wavefunction can be represented
according to the Berry conjecture \cite{berry1977}. In the Srednicki
form \cite{srednicki1994}, the non-normalized solution of \prettyref{eq:SchrPhi},
labeled by the index $n$, with the well-defined total spin $S$ is
a superposition of plane waves
\begin{multline}
\Phi_{tn}^{(S)}\propto\sum_{\{\mathbf{p}\}}A_{n}^{(S)}(t,\{\mathbf{p}\})\tilde{\delta}(\{\mathbf{p}\}^{2}-2mE_{n}^{(S)})\\
\times\exp(i\sum_{j}\mathbf{p}_{j}\mathbf{r}_{j}/\hbar)\label{eq:Phiexp}
\end{multline}
with the momenta $\mathbf{p}_{j}$ in the periodic box with incommensurable
dimensions. Due to discrete spectrum of $\mathbf{p}_{j}$, the states
with approximately fixed energies $E_{n}^{(S)}$ are selected by the
function
\[
\tilde{\delta}(x)=\Theta(\varDelta-|x|)/(2\varDelta),
\]
where $\Theta(x)$ is the Heaviside step function. In the final calculations,
when the summation over $\{\mathbf{p}\}$ is replaced by integration,
$\tilde{\delta}(x)$ is replaced by the Dirac $\delta$-function.
Then the total wavefunction \eqref{eq:PsiPhiXi} can be represented
as (see Appendix \ref{sec:Wavefunctions-Int-NonInt}) 
\begin{equation}
\Psi_{nS_{z}}^{(S)}=\mathcal{N}_{n}^{(S)}\sum_{r}{\sum_{\{\mathbf{p}\}}}'A_{n}^{(S)}(r,\{\mathbf{p}\})\tilde{\delta}(\{\mathbf{p}\}^{2}-2mE_{n}^{(S)})\tilde{\Psi}_{r\{\mathbf{p}\}S_{z}}^{(S)}.\label{eq:PsiSnSz}
\end{equation}
It is a superposition of symmetrized plane waves --- wavefunctions
$\tilde{\Psi}_{r\{\mathbf{p}\}S_{z}}^{(S)}$ of non-interacting particles
(see \prettyref{eq:tilPsi} and \cite{kaplan,pauncz_symmetric,yurovsky2015}).
Given $S$ and $S_{z}$, these wavefunctions are labeled by the Young
tableau $r$ and the set of particle momenta $\{\mathbf{p}\}\equiv\{\mathbf{p}_{1},\ldots,\mathbf{p}_{N}\}$
($\{\mathbf{p}\}^{2}\equiv\sum_{j}\mathbf{p}_{j}^{2}$). The summation
over the simplex $\mathbf{p}_{1}<\mathbf{p}_{2}<\cdots<\mathbf{p}_{N}$
is denoted as $\sum'$, where $\mathbf{p}<\mathbf{p'}$ if $p_{x}<p'_{x}$,
or $p_{x}=p'_{x}$ and $p_{y}<p'_{y}$, or $p_{x}=p'_{x}$, $p_{y}=p'_{y}$,
and $p_{z}<p'_{z}$. Such summation, neglecting multiple occupations
of the momentum states, is applicable to non-degenerate gases, when
the difference between Bose-Einstein, Fermi-Dirac, and Boltzmann distributions
is negligibly small. The normalization factor (see Appendix \ref{sec:Normalization})
$\mathcal{N}_{n}^{(S)}$ provides $\langle\Psi_{nS_{z}}^{(S)}|\Psi_{nS_{z}}^{(S)}\rangle=1$.

According to the Berry's conjecture \cite{berry1977,srednicki1994},
the coefficients $A_{n}^{(S)}(r,\{\mathbf{p}\})$ are treated as Gaussian
random variables with a two-point correlation function (see Appendix
\ref{sec:Correlation}), generalizing the one of \cite{srednicki1994}
to the states with well-defined spins, 
\begin{equation}
\left\langle {A_{n'}^{(S')}}^{*}(r',\{\mathbf{p'}\})A_{n}^{(S)}(r,\{\mathbf{p}\})\right\rangle _{\mathrm{EE}}=\frac{\delta_{S'S}\delta_{n'n}\delta_{r'r}\delta_{\{\mathbf{p'}\}\{\mathbf{p}\}}}{\tilde{\delta}(\{\mathbf{p'}\}^{2}-\{\mathbf{p}\}^{2})}.\label{eq:TwoPointCorr}
\end{equation}
Here, as in \cite{srednicki1994}, $\left\langle \right\rangle _{\mathrm{EE}}$
denotes average over a fictitious ``eigenstate ensemble'', which
describes properties of a typical eigenfunction. The Kronecker symbols
appear here, as well as in the correlation function \cite{srednicki1994},
since different $S$ and $n$ correspond to different (independent)
eigenfunctions and different $\{\mathbf{p}\}$ in the same simplex
correspond to different (independent) plane waves. In addition, \prettyref{eq:TwoPointCorr}
contains the Kronecker symbol of the Young tableaux $r$ and $r'$,
as proved in Appendix \ref{sec:Correlation}.

\section{Decay rates\label{sec:Decay}}

The rate of transitions from the state with the spin $S$ to the $S'$
one is estimated by the Weisskopf-Wigner width (see \cite{agarwal})
\begin{equation}
\Gamma_{S_{z}}^{(S,S')}=\frac{2\pi}{\hbar}|\langle\Psi_{n'S'_{z}}^{(S')}|\hat{V}_{\mathrm{spin}}|\Psi_{nS_{z}}^{(S)}\rangle|^{2}\frac{dn^{(S')}(E_{n'}^{(S')})}{dE}|_{E_{n'}^{(S')}=E_{n}^{(S)}},\label{eq:GammaWW}
\end{equation}
where the density of states $dn^{(S)}/dE$ is evaluated in Appendix
\ref{sec:Normalization}. For a typical wavefunction \eqref{eq:PsiSnSz},
the squared modulus of the matrix element can be estimated by the
eigenstate-ensemble average 
\begin{multline}
\left\langle |\langle\Psi_{n'S'_{z}}^{(S')}|\hat{V}_{\mathrm{spin}}|\Psi_{nS_{z}}^{(S)}\rangle|^{2}\right\rangle _{\mathrm{EE}}=(\mathcal{N}_{n}^{(S)}\mathcal{N}_{n'}^{(S')})^{2}\\
\times{\sum_{\{\mathbf{p}\}}}'{\sum_{\{\mathbf{p'}\}}}'\tilde{\delta}(\{\mathbf{p}\}^{2}-2mE_{n}^{(S)})\tilde{\delta}(\{\mathbf{p'}\}^{2}-2mE_{n'}^{(S')})\\
\times\sum_{r,r'}|\langle\tilde{\Psi}_{r'\{\mathbf{p}'\}S'_{z}}^{(S')}|\hat{V}_{\mathrm{spin}}|\tilde{\Psi}_{r\{\mathbf{p}\}S_{z}}^{(S)}\rangle|^{2}.\label{eq:MatElPsi2}
\end{multline}
Here $S'\neq S$ and the product of four coefficients $A_{n}^{(S)}(r,\{\mathbf{p}\})$
leads to a four-point correlation function \eqref{eq:FourPointCorr},
which is represented by \prettyref{eq:FourTwoCorr} in terms of two-point
ones \eqref{eq:TwoPointCorr} for the Gaussian ensemble as in \cite{srednicki1994}.
This relation of matrix elements between wavefunctions of interacting
and non-interacting particles is obtained since the correlation function
\eqref{eq:TwoPointCorr} contains $\delta_{r'r}$. The sum of squared
moduli of the matrix elements between the wavefunctions of non-interacting
particles in \prettyref{eq:MatElPsi2} is calculated with the sum
rules \cite{yurovsky2015a}. 

The expansion \eqref{eq:VspinSph} of the interactions $\hat{V}_{\mathrm{spin}}$
contains irreducible spherical scalar, vector, and tensor. According
to the Wigner-Eckart theorem, the scalar interaction $\hat{V}$ conserves
the total spin, while the spins $S$ and $S'$ can be coupled by the
vector component $\hat{V}_{0}$ if $|S-S'|\leq1$ and by the rank
2 tensor component $\hat{V}_{0}^{(2)}$ if $|S-S'|\leq2$. This leads
to quantum interference of the vector and tensor contributions to
the transitions between states with spins $S$ and $S'=S\pm1$. Equation
\eqref{eq:MatElPsi2} and the sum rules \cite{yurovsky2015a} lead
(see Appendix \ref{sec:Matrix-elements}) to the transition rates
\begin{subequations}\label{GammaSSz} 
\begin{multline}
\Gamma_{S_{z}}^{(S,S-1)}=\frac{(S^{2}-S_{z}^{2})(N+2S+2)}{S(2S+1)N}\\
\times\left[\alpha_{+}^{2}\frac{S_{z}^{2}(N+2)}{S^{2}-1}+\alpha_{-}^{2}(N-2)+4\alpha_{+}\alpha_{-}S_{z}\right]\Gamma_{Dd}
\end{multline}
\begin{multline}
\Gamma_{S_{z}}^{(S,S+1)}=\frac{[(S+1)^{2}-S_{z}^{2}](N-2S)}{(S+1)(2S+1)N}\\
\times\left[\alpha_{+}^{2}\frac{S_{z}^{2}(N+2)}{S(S+2)}+\alpha_{-}^{2}(N-2)+4\alpha_{+}\alpha_{-}S_{z}\right]\Gamma_{Dd}
\end{multline}
\begin{multline}
\Gamma_{S_{z}}^{(S,S-2)}=\frac{[(S-1)^{2}-S_{z}^{2}](S^{2}-S_{z}^{2})}{2S(S-1)(2S-1)(2S+1)N}\\
\times(N+2S)(N+2S+2)\alpha_{+}^{2}\Gamma_{Dd}
\end{multline}
\begin{multline}
\Gamma_{S_{z}}^{(S,S+2)}=\frac{[(S+1)^{2}-S_{z}^{2}][(S+2)^{2}-S_{z}^{2}]}{2(S+1)(S+2)(2S+1)(2S+3)N}\\
\times(N-2S)(N-2S-2)\alpha_{+}^{2}\Gamma_{Dd}
\end{multline}
\end{subequations}where $\alpha_{+}=(a_{\upuparrows}+a_{\downdownarrows}-2a_{\uparrow\downarrow})/a_{S}$,
$\alpha_{-}=(a_{\upuparrows}-a_{\downdownarrows})/a_{S}$, $a_{S}$
is given by \prettyref{eq:aSchaos3d}, and the interference terms
are proportional to $\alpha_{+}\alpha_{-}$. 

The rates are proportional to the factors (see Appendix \ref{sec:Matrix-elements})
\begin{equation}
\Gamma_{3d}=2\sqrt{\pi T/m}a_{S}^{2}n_{3d},\quad\Gamma_{2d}=\frac{\pi}{2}a_{S}^{2}\omega_{\mathrm{conf}}n_{2d}\label{eq:GammaDd}
\end{equation}
where $n_{Dd}$ is the $D$-dimensional gas density and the temperature
is related to the eigenstate energy as $T=2E_{n}^{(S)}/(3N)$ \cite{srednicki1994}.
In the 3D case ($D=3$), $\Gamma_{3d}$, up to a numerical factor,
is the frequency of elastic collisions per particle in the gas. In
the 2D case, $\Gamma_{2d}$ is proportional to the rate of collisions
per particle too, since the probability of collision during one axial
oscillation is $8\pi a_{S}^{2}n_{2d}$, the oscillation frequency
is $2\pi\omega_{\mathrm{conf}}$, and the oscillation velocity substantially
exceeds the 2D motion one in the 2D regime ($T<\hbar\omega_{\mathrm{conf}}$).
The present derivation is valid whenever $T$ substantially exceeds
the degeneracy temperature $T_{\mathrm{deg}}=2\pi\hbar^{2}n_{3d}^{2/3}/m$
\cite{pitaevskii}.

For the two states of $^{87}$Rb atoms , generally used in experiments,
$|\downarrow\rangle=|F=1,m_{f}=1\rangle$ and $|\uparrow\rangle=|F=2,m_{f}=-1\rangle$,
the scattering lengths are \cite{egorov2011} $a_{\downdownarrows}\approx100.4a_{B}$,
$a_{\upuparrows}\approx95.5a_{B}$, and $a_{\uparrow\downarrow}\approx98.0a_{B}$,
where $a_{B}$ is the Bohr radius. For $n_{3d}=10^{12}\mathrm{cm}^{-3}$
we have $T_{\mathrm{deg}}\approx0.04\mathrm{\mu K}$. The zero-range
approximation \eqref{eq:ga3d2d} is applicable whenever $T\ll10^{4}\mu\mathrm{K}$
and, according to the criterion \eqref{eq:aSchaos3d}, the system
of $N=10^{4}$ atoms becomes chaotic at $T>8\times10^{-13}\mu\mathrm{K}$.
Then for $T=1\mu\mathrm{K}$ we have $\Gamma_{3d}\approx0.9\mathrm{s}^{-1}$. 

\begin{figure}
\includegraphics[width=3.4in]{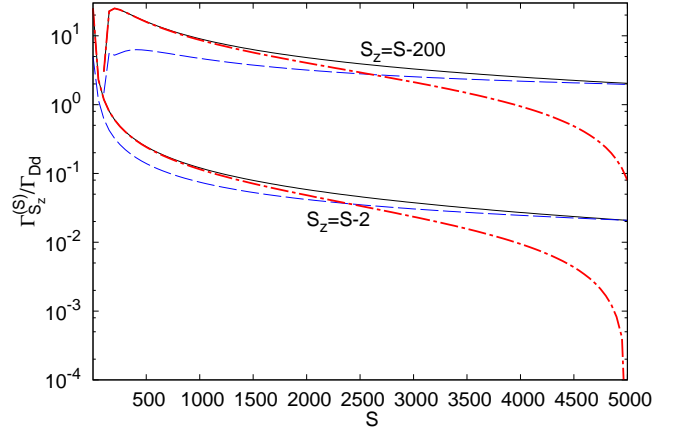}

\caption{The total decay rates for the state with the total spin $S$ and its
projection $S_{z}$ for two values of $S-S_{z}$. The solid black
lines correspond to the background scattering lengths. The dashed
blue and dot-dashed red lines show the decay rates minimized by tuning
of $a_{\downdownarrows}$ and $a_{\uparrow\downarrow}$, respectively.\label{Fig_gam_dd_ud}}

\end{figure}

\begin{figure}
\includegraphics[width=3.4in]{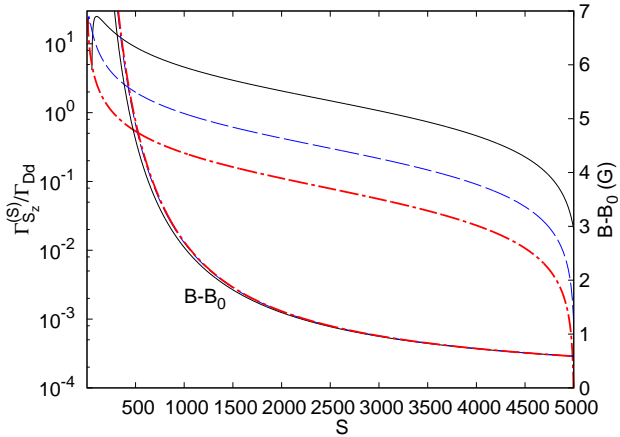}

\caption{The total decay rates for the state with the total spin $S$ and its
projection $S_{z}$ minimized by $a_{\uparrow\downarrow}$ tuning
in the vicinity of the Feshbach resonance at $B_{0}\approx9.13\mathrm{G}$
in $^{87}$Rb. The solid black, dashed blue, and dot-dashed red lines
correspond to $S-S_{z}=100$, 20, and 5, respectively. The plots of
optimal detuning $B-B_{0}$ for different $S-S_{z}$ almost coincide.\label{Fig_gaud_db}}

\end{figure}

\begin{figure}
\includegraphics[width=3.4in]{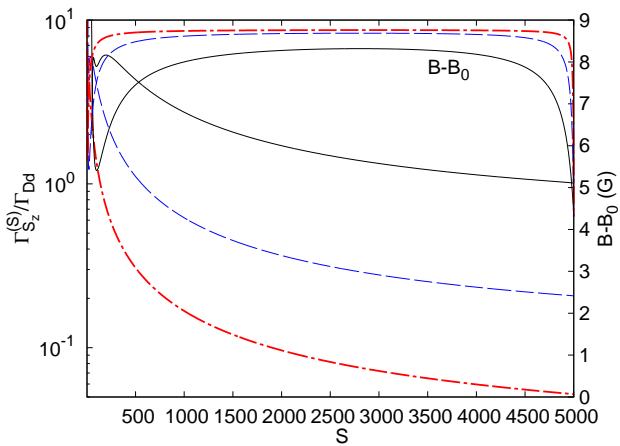}

\caption{The total decay rates for the state of $N=10^{4}$ atoms of $^{87}$Rb
with the total spin $S$ and its projection $S_{z}$ minimized by
$a_{\downdownarrows}$ tuning in the vicinity of the Feshbach resonance
at $B_{0}\approx1007.4\mathrm{G}$. The solid black, dashed blue,
and dot-dashed red lines correspond to $S-S_{z}=100$, 20, and 5,
respectively. The optimal detuning $B-B_{0}$ for different $S-S_{z}$
is presented by three upper plots.\label{Fig_gadd_db}}
\end{figure}

The total decay rate $\Gamma_{S_{z}}^{(S)}=\sum_{S'\neq S}\Gamma_{S_{z}}^{(S,S')}$
is presented in Fig. \ref{Fig_gam_dd_ud}. Even for $N=10^{4}$ atoms
it does not exceed $\approx60\Gamma_{Dd}$. The decay is suppressed
at large values of $S$ and $S_{z}$.

The Feshbach resonance tuning of the elastic scattering length \cite{chin2010}
cannot eliminate the decay, as one tuning parameter --- the magnetic
field --- cannot make vanish two combination of the scattering lengths
--- $\alpha_{+}$ and $\alpha_{-}$. However, the Feshbach tuning
can minimize the decay due to the destructive interference of the
contributions of the spherical vector and tensor interactions, mentioned
above. These contributions are proportional to $\alpha_{-}$ and $\alpha_{+}$,
respectively, in \refeq{GammaSSz}. In the case of $^{87}$Rb, the
resonance in $\uparrow\downarrow$ collision at $B_{0}\approx9.13\mathrm{G}$
(with the width $\Delta\approx15\mathrm{mG}$) is well separated from
the resonances in $\downarrow\downarrow$ collisions at $B_{0}>400\mathrm{G}$
(the widest one at $B_{0}\approx1007.4\mathrm{G}$ has $\Delta\approx0.21\mathrm{G}$)
\cite{chin2010}. No resonances are known in $\uparrow\uparrow$ collisions.
Then either $a_{\uparrow\downarrow}$ or $a_{\downdownarrows}$ can
be tuned without changing other scattering lengths. The resulting
decay rates are presented in Fig. \ref{Fig_gam_dd_ud}. Tuning of
$a_{\downdownarrows}$ can reduce the decay rate at small $S$, while
tuning of $a_{\uparrow\downarrow}$ can lead to the reduction by orders
of magnitude at large $S$ and $S_{z}$. The magnetic field detunings,
minimizing the decay rates, and the minimal rates are plotted in Figs.
\ref{Fig_gaud_db} and \ref{Fig_gadd_db}. Even the minimal detunings
$0.5\mathrm{G}$ and $4\mathrm{G}$, respectively, substantially exceeds
the resonance widths. Then the closed-channel effects can be neglected,
although the resonances are closed-channel dominated \cite{chin2010}.
The resonance-enhancement of three-body losses (see \cite{yurovsky2003})
can be neglected at such detunings as well. 

Approximate expressions can be obtained in the regions of maximal
suppression, $S_{z}\approx\pm S$. Here the $a_{\uparrow\downarrow}$
tuning minimizes the decay rate when $\alpha_{+}\approx\mp\alpha_{-}$
and the minimal rate is approximated for $N\gg1$ by 
\begin{multline*}
\Gamma_{S_{z}}^{(S)}\approx\frac{2\alpha_{+}^{2}\Gamma_{Dd}}{N(2S+3)}\biggl\{2(N-2S)(N-2S-2)\\
-\frac{S-|S_{z}|}{S+1}\left[2N^{2}-(N-2S)N(4S+5)\right]\biggr\}.
\end{multline*}
Then for $N/2-S=\mathrm{const}\ll N$, $S-|S_{z}|=\mathrm{const}$,
and fixed density the decay rate is scaled as $1/N$. The desired
scattering lengths, $a_{\uparrow\downarrow}\approx a_{\upuparrows}$
at $S_{z}\approx S$, or $a_{\uparrow\downarrow}\approx a_{\downdownarrows}$
at $S_{z}\approx-S$ , are obtained for $^{87}$Rb at $B-B_{0}\approx0.6G$
or $-0.6G$, respectively. The state with $S=S_{z}=N/2$ for even
$N$ (or with $S=S_{z}=(N-1)/2$ for odd $N$ ) cannot decay to states
with $S'<S$ since $S'$ cannot be less than $S_{z}$, nor to states
with $S'>S$ since $S'$ cannot exceed $N/2$.

When the decay rates are substantially suppressed, the lifetime of
the states with well-defined spins is restricted by the loss processes
in the cold gas, such as spin-changing (dipolar relaxation) and three-body
collisions, leading to high-energy atoms escaping the trap. The spin-changing
collisions can also change many-body spin and its projection, leading
to additional decay. However, this decay, having the rate comparable
to the gas loss rate, does not lead to additional restriction of the
lifetime of the states with well-defined spins. In real physical situations,
dipolar relaxation becomes substantial only for Cs and atoms with
high magnetic momenta \cite{stamper2013}.

\section*{Conclusions}

Spatially-chaotic many-body eigenstates of the total spin operator
can be described, according to the Berry conjecture \cite{berry1977},
within the Srednicki approach \cite{srednicki1994} {[}see \prettyref{eq:PsiSnSz}{]}.
This description, in a combination with the sum rules \cite{yurovsky2015a},
allows us to evaluate the matrix elements of spin-dependent two-body
interactions, leading to transitions between states with different
total spins. The transition rates, calculated within the Weisskopf-Wigner
approach \refneq{GammaSSz}, are proportional to the elastic collision
rate per particle \eqref{eq:GammaDd}. The decay rates can be suppressed
due to destructive interference of the contributions the spherical
vector and tensor terms in the spin-dependent interaction $\hat{V}_{\mathrm{spin}}$.
The interference terms in \refeq{GammaSSz} can be controlled by Feshbach
resonances as they are proportional to $\alpha_{+}\alpha_{-}$. Another
manifestation of quantum interference is the effect of dynamical localization.
It can slow down relaxation due to periodic driving \cite{das2010,roy2015,cemkeser2015},
while the present mechanism pertains to time-independent systems.
The long-lived entangled states can find applications in quantum computation
and metrology.

\appendix

\section{Wavefunctions with defined total spins of interacting and non-interacting
particles\label{sec:Wavefunctions-Int-NonInt}}

A permutation of coordinates in the wavefunction \eqref{eq:Phiexp}
can be represented as
\begin{multline*}
\pr P\Phi_{tn}^{(S)}\propto\sum_{\{\mathbf{p}\}}A_{n}^{(S)}(t,\{\mathbf{p}\})\tilde{\delta}(\{\mathbf{p}\}^{2}-2mE_{n}^{(S)})\\
\times\exp(i\sum_{j}\mathbf{p}_{\pr P^{-1}j}\mathbf{r}_{j}/\hbar)\\
\propto\sum_{\{\mathbf{p}\}}A_{n}^{(S)}(t,\pr P\{\mathbf{p}\})\tilde{\delta}(\{\mathbf{p}\}^{2}-2mE_{n}^{(S)})\\
\times\exp(i\sum_{j}\mathbf{p}_{j}\mathbf{r}_{j}/\hbar),
\end{multline*}
where $\pr P\{\mathbf{p}\}\equiv\{\mathbf{p}_{\pr P1},\ldots,\mathbf{p}_{\pr PN}\}$.
Then \prettyref{eq:PermPhiXi} leads to
\begin{equation}
A_{n}^{(S)}(t,\pr P\{\mathbf{p}\})=\sum_{r}D_{rt}^{[\lambda]}(\pr P)A_{n}^{(S)}(r,\{\mathbf{p}\})\label{eq:PermA}
\end{equation}

Let us represent the wavefunction $\Phi_{tn}^{(S)}$ in the form which
explicitly shows its permutation properties. This can be done by summation
in \prettyref{eq:Phiexp} over the simplex $\mathbf{p}_{1}<\mathbf{p}_{2}<\cdots<\mathbf{p}_{N}$.
Momentum sets in other simplices are given by $\pr P\{\mathbf{p}\}$.
Neglecting contributions of the sets $\{\mathbf{p}\}$ which contain
equal momenta $\mathbf{p}_{j}=\mathbf{p}_{j'}$, one gets
\begin{multline}
\Phi_{tn}^{(S)}\propto\sum_{\pr P}{\sum_{\{\mathbf{p}\}}}'A_{n}^{(S)}(t,\pr P\{\mathbf{p}\})\tilde{\delta}(\{\mathbf{p}\}^{2}-2mE_{n}^{(S)})\\
\times\exp(i\sum_{j}\mathbf{p}_{\pr Pj}\mathbf{r}_{j}/\hbar)\\
\propto{\sum_{\{\mathbf{p}\}}}'\sum_{r}A_{n}^{(S)}(r,\{\mathbf{p}\})\tilde{\delta}(\{\mathbf{p}\}^{2}-2mE_{n}^{(S)})\sum_{\pr P}D_{rt}^{[\lambda]}(\pr P)\\
\times\exp(i\sum_{j}\mathbf{p}_{j}\mathbf{r}_{\pr P^{-1}j}/\hbar)\\
\propto{\sum_{\{\mathbf{p}\}}}'\sum_{r}A_{n}^{(S)}(r,\{\mathbf{p}\})\tilde{\delta}(\{\mathbf{p}\}^{2}-2mE_{n}^{(S)})\left(\frac{N!}{f_{S}}\right)^{1/2}\tilde{\Phi}_{tr\{\mathbf{p}\}}^{(S)}.\label{eq:PhitilPhi}
\end{multline}
Here \prettyref{eq:PermA} and the identity for the orthogonal representation
matrices $D_{rt}^{[\lambda]}(\pr P^{-1})=D_{tr}^{[\lambda]}(\pr P)$
are used and
\[
\tilde{\Phi}_{tr\{\mathbf{p}\}}^{(S)}=\left(\frac{f_{S}}{N!}\right)^{1/2}\sum_{\pr P}D_{tr}^{[\lambda]}(\pr P)\exp(i\sum_{j}\mathbf{p}_{j}\mathbf{r}_{\pr Pj}/\hbar)
\]
are spatial wavefunctions of non-interacting particles \cite{kaplan,pauncz_symmetric,yurovsky2015}.
They satisfy relations \eqref{eq:PermPhiXi}, forming a basis of the
irreducible representation associated with $\lambda$. The Young tableau
$r$ labels different representations for the same $S$ and $\{\mathbf{p}\}$.
The total wavefunction of non-interacting particles 
\begin{equation}
\tilde{\Psi}_{r\{\mathbf{p}\}S_{z}}^{(S)}=f_{S}^{-1/2}\sum_{t}\tilde{\Phi}_{tr\{\mathbf{p}\}}^{(S)}\Xi_{tS_{z}}^{(S)}\label{eq:tilPsi}
\end{equation}
is expressed similarly to $\Psi_{nS_{z}}^{(S)}$ {[}see \prettyref{eq:PsiPhiXi}{]}.
Then \prettyref{eq:PhitilPhi} leads to \prettyref{eq:PsiSnSz}. The
wavefunctions of non-interacting particles form an orthonormal basis
\begin{equation}
\left\langle \tilde{\Psi}_{r'\{\mathbf{p'}\}S'_{z}}^{(S')}|\tilde{\Psi}_{r\{\mathbf{p}\}S_{z}}^{(S)}\right\rangle =\delta_{S'S}\delta_{S'_{z}S_{z}}\delta_{r'r}\delta_{\{\mathbf{p'}\}\{\mathbf{p}\}}\label{eq:tilPsiorth}
\end{equation}
and satisfy the Schr\"odinger equation 
\[
\hat{H}_{0}\tilde{\Psi}_{r\{\mathbf{p}\}S_{z}}^{(S)}=\frac{1}{2m}\{\mathbf{p}\}^{2}\tilde{\Psi}_{r\{\mathbf{p}\}S_{z}}^{(S)}.
\]

\section{Two- and four-point correlation functions of the coefficients $A_{n}^{(S)}(r,\{\mathbf{p}\})$\label{sec:Correlation}}

According to the Berry conjecture \cite{berry1977}, the coefficients
$A_{n}^{(S)}(t,\{\mathbf{p}\})$ in Eqs. \eqref{eq:Phiexp} and \eqref{eq:PsiSnSz}
can be treated as Gaussian random variables with a two-point correlation
function \cite{srednicki1994}
\begin{equation}
\left\langle {A_{n'}^{(S')}}^{*}(t',\{\mathbf{p'}\})A_{n}^{(S)}(t,\{\mathbf{p}\})\right\rangle _{\mathrm{EE}}=\frac{\delta_{S'S}\delta_{n'n}\delta_{\{\mathbf{p'}\}\{\mathbf{p}\}}f(t,t')}{\tilde{\delta}(\{\mathbf{p'}\}^{2}-\{\mathbf{p}\}^{2})},\label{eq:TwoPointCorrf}
\end{equation}
where ${A_{n}^{(S)}}^{*}(t,\{\mathbf{p}\})=A_{n}^{(S)}(t,\{\mathbf{-p}\})$.
The Kronecker symbols appear here since different $S$ and $n$ correspond
to different (independent) eigenfunctions and different $\{\mathbf{p}\}$
within the given simplex correspond to different (independent) plane
waves. (The correlation functions with $\{\mathbf{p}\}$ and $\{\mathbf{p'}\}$
in different simplices do not appear within the present paper.) By
now nothing can be told on $f(t,t')$, since $\Phi_{tn}^{(S)}$ and
$\Phi_{t'n}^{(S)}$ are components of the same eigenfunction, related
by \prettyref{eq:PermPhiXi}. The factors $f(t,t')$ are determined
below.

Equation \eqref{eq:PermA} leads to the following equality for an
arbitrary permutation $\pr P$ 
\[
A_{n}^{(S)}(t,\{\mathbf{p}\})=\sum_{r}D_{tr}^{[\lambda]}(\pr P)A_{n}^{(S)}(r,\pr P\{\mathbf{p}\}).
\]
It relates coefficients $A_{n}^{(S)}(r,\{\mathbf{p}\})$ in different
simplices and leads to
\begin{multline*}
\left\langle {A_{n}^{(S)}}^{*}(t',\{\mathbf{p}\})A_{n}^{(S)}(t,\{\mathbf{p}\})\right\rangle _{\mathrm{EE}}\\
=\sum_{r,r'}D_{t'r'}^{[\lambda]}(\pr P)D_{tr}^{[\lambda]}(\pr P)\left\langle {A_{n}^{(S)}}^{*}(r',\pr P\{\mathbf{p}\})A_{n}^{(S)}(r,\pr P\{\mathbf{p}\})\right\rangle _{\mathrm{EE}}.
\end{multline*}
 Applying \prettyref{eq:TwoPointCorrf} to all correlation functions,
we get 
\begin{equation}
f(t,t')=\sum_{r,r'}D_{t'r'}^{[\lambda]}(\pr P)D_{tr}^{[\lambda]}(\pr P)f(r,r')\label{eq:eqfttp}
\end{equation}
for any $\pr P$. Averaging the right hand side of this equation over
all $\pr P$ and using the orthogonality relation\cite{kaplan,pauncz_symmetric}
\[
\sum_{\pr P}D_{t'r'}^{[\lambda']}(\pr P)D_{tr}^{[\lambda]}(\pr P)=\frac{N!}{f_{S}}\delta_{tt'}\delta_{rr'}\delta_{\lambda\lambda'},
\]
one gets the equation
\[
f(t,t')=\frac{1}{f_{S}}\delta_{tt'}\sum_{r}f(r,r).
\]
Its solution is $f(t,t')=\mathrm{const}\delta_{tt'}$. Equation \eqref{eq:TwoPointCorr}
is obtained if $\mathrm{const}=1$. Other choice of the constant factor
can only change the normalization factor in \prettyref{eq:PsiSnSz}.
The function $f(t,t')=\delta_{tt'}$ satisfies \prettyref{eq:eqfttp},
as can be easily proven using the identity
\begin{equation}
\sum_{r}D_{t'r}^{[\lambda]}(\pr P)D_{tr}^{[\lambda]}(\pr P)=D_{t't}^{[\lambda]}(\pr P\pr P^{-1})=\delta_{t't}\label{eq:ProdInv}
\end{equation}
for the Young orthogonal matrices (see \cite{kaplan,pauncz_symmetric}). 

The relation between four-point and two-point correlation functions\begin{widetext}
\begin{multline}
\left\langle A_{n'''}^{(S''')}(t''',\{\mathbf{p'''}\})A_{n''}^{(S'')}(t'',\{\mathbf{p''}\})A_{n'}^{(S')}(t',\{\mathbf{p'}\})A_{n}^{(S)}(t,\{\mathbf{p}\})\right\rangle _{\mathrm{EE}}\\
=\left\langle A_{n'''}^{(S''')}(t''',\{\mathbf{p'''}\})A_{n''}^{(S'')}(t'',\{\mathbf{p''}\})\right\rangle _{\mathrm{EE}}\left\langle A_{n'}^{(S')}(t',\{\mathbf{p'}\})A_{n}^{(S)}(t,\{\mathbf{p}\})\right\rangle _{\mathrm{EE}}\\
+\left\langle A_{n'''}^{(S''')}(t''',\{\mathbf{p'''}\})A_{n'}^{(S')}(t',\{\mathbf{p'}\})\right\rangle _{\mathrm{EE}}\left\langle A_{n''}^{(S'')}(t'',\{\mathbf{p''}\})A_{n}^{(S)}(t,\{\mathbf{p}\})\right\rangle _{\mathrm{EE}}\\
+\left\langle A_{n''}^{(S'')}(t'',\{\mathbf{p''}\})A_{n'}^{(S')}(t',\{\mathbf{p'}\})\right\rangle _{\mathrm{EE}}\left\langle A_{n'''}^{(S''')}(t''',\{\mathbf{p'''}\})A_{n}^{(S)}(t,\{\mathbf{p}\})\right\rangle _{\mathrm{EE}}\label{eq:FourTwoCorr}
\end{multline}
is a straightforward generalization of the similar relation in \cite{srednicki1994}.
For the product of four coefficients $A_{n}^{(S)}(r,\{\mathbf{p}\})$
in \prettyref{eq:MatElPsi2} it leads to
\begin{equation}
\left\langle {A_{n'}^{(S')}}^{*}(r',\{\mathbf{p'}\})A_{n}^{(S)}(r,\{\mathbf{p}\}){A_{n}^{(S)}}^{*}(r'',\{\mathbf{p''}\})A_{n'}^{(S')}(r''',\{\mathbf{p'''}\})\right\rangle _{\mathrm{EE}}=\frac{\delta_{r''r}\delta_{\{\mathbf{p''}\}\{\mathbf{p}\}}\delta_{r'''r'}\delta_{\{\mathbf{p'''}\}\{\mathbf{p'}\}}}{\tilde{\delta}(\{\mathbf{p''}\}^{2}-\{\mathbf{p}\}^{2})\tilde{\delta}(\{\mathbf{p'''}\}^{2}-\{\mathbf{p'}\}^{2})}\label{eq:FourPointCorr}
\end{equation}
\end{widetext}since $S'\neq S$.

\section{Normalization factor and the density of states\label{sec:Normalization}}

Orthonormality of the non-interacting particle wavefunctions \eqref{eq:tilPsiorth}
leads, using the two-point correlation functions \eqref{eq:TwoPointCorr},
to the following overlap of the wavefunctions of interacting particles
\eqref{eq:PsiSnSz}
\begin{multline}
\left\langle \left\langle \Psi_{n'S'_{z}}^{(S')}|\Psi_{nS_{z}}^{(S)}\right\rangle \right\rangle _{\mathrm{EE}}=\left(\mathcal{N}_{n}^{(S)}\right)^{2}f_{S}\delta_{S'_{z}S_{z}}\delta_{S'S}\delta_{n'n}\\
\times{\sum_{\{\mathbf{p}\}}}'\tilde{\delta}(\{\mathbf{p}\}^{2}-2mE_{n}^{(S)}).\label{eq:overPsi}
\end{multline}
The summation over $\{\mathbf{p}\}$ in the simplex can be approximated
by integration over whole momentum space with the replacement of $\tilde{\delta}$
by the Dirac $\delta$-function,
\begin{multline}
{\sum_{\{\mathbf{p}\}}}'\tilde{\delta}(\{\mathbf{p}\}^{2}-2mE_{n}^{(S)})\approx\frac{1}{N!}\left(\frac{L}{2\pi\hbar}\right)^{DN}\\
\times\int d^{DN}p\delta(\{\mathbf{p}\}^{2}-2mE_{n}^{(S)})\label{eq:SumInt}
\end{multline}
The integral
\begin{equation}
\int d^{DN}p\delta(\{\mathbf{p}\}^{2}-E)=\frac{(\pi E)^{DN/2}}{\Gamma(DN/2)E}\label{eq:IntDelta}
\end{equation}
is calculated in \cite{srednicki1994}. Then the normalization factor
is given by
\begin{equation}
\left(\mathcal{N}_{n}^{(S)}\right)^{-2}=\frac{f_{S}L^{DN}(mE_{n}^{(S)})^{DN/2-1}}{2N!(2\pi)^{DN/2}\Gamma(DN/2)\hbar^{DN}}.\label{eq:NormFact}
\end{equation}

The density of states of interacting particles can be approximated
by the one of non-interacting ones. Number of such states with the
total spin $S$ below the energy $E$ is
\[
n^{(S)}(E)=f_{S}{\sum_{\{\mathbf{p}\}}}'\Theta(E-\{\mathbf{p}\}^{2}/(2m)).
\]
Then the density of states is estimated as 
\begin{multline}
\frac{dn^{(S)}(E)}{dE}\approx\frac{n^{(S)}(E+\varDelta)-n^{(S)}(E-\varDelta)}{2\varDelta}\\
=f_{S}{\sum_{\{\mathbf{p}\}}}'\tilde{\delta}(E-\{\mathbf{p}\}^{2}/(2m))=2m\left(\mathcal{N}_{n}^{(S)}\right)^{-2}\label{eq:DenStatesN}
\end{multline}
{[}cf. \prettyref{eq:overPsi}{]}.

\section{Matrix elements and decay rates\label{sec:Matrix-elements}}

If the scattering lengths $a_{\sigma\sigma'}$ and interaction strengths
$g_{Dd}^{\sigma\sigma'}$ are spin-dependent, states with different
total spins become coupled by the spherical vector $\hat{V}_{0}$
and tensor $\hat{V}_{0}^{(2)}$ interactions in \prettyref{eq:VspinSph}.
Matrix elements of the spherical vector and tensor are related to
ones for the maximal allowed spin projections by the Wigner-Eckart
theorem \cite{yurovsky2015,yurovsky2015a},
\begin{equation}
\begin{aligned}\langle\Psi_{n'S_{z}}^{(S')}|\hat{V}_{0}|\Psi_{nS_{z}}^{(S)}\rangle= & X_{S_{z}0}^{(S,S',1)}\langle\Psi_{n'S'}^{(S')}|\hat{V}_{S'-S}|\Psi_{nS}^{(S)}\rangle\\
\langle\Psi_{n'S_{z}}^{(S')}|\hat{V}_{0}^{(2)}|\Psi_{nS_{z}}^{(S)}\rangle= & X_{S_{z}0}^{(S,S',2)}\langle\Psi_{n'S'}^{(S')}|\hat{V}_{S'-S}^{(2)}|\Psi_{nS}^{(S)}\rangle,
\end{aligned}
\label{eq:WigEcc}
\end{equation}
and do not vanish for $|S-S'|\leq1$ and $|S-S'|\leq2$, respectively.
Here the ratios of the $3j$ Wigner symbols $X_{S_{z}0}^{(S,S',q)}$
are tabulated in \cite{yurovsky2015,yurovsky2015a}. Matrix elements
with $S'>S$ are calculated using Hermitian conjugate in \prettyref{eq:WigEcc},
taking into account that $(\hat{V}_{0})^{\dagger}=\hat{V}_{0}$, $(\hat{V}_{0}^{(2)})^{\dagger}=\hat{V}_{0}^{(2)}$,
$\hat{V}_{-1}^{\dagger}=-\hat{V}_{+1}$, $(\hat{V}_{-1}^{(2)})^{\dagger}=-\hat{V}_{+1}^{(2)}$,
and $(\hat{V}_{-2}^{(2)})^{\dagger}=\hat{V}_{+2}^{(2)}$ (see \cite{yurovsky2015a}).

Sums of the products of the matrix elements of the spherical tensors
between the wavefunctions of non-interacting particles \eqref{eq:tilPsi}
can be represented for $S'\leq S$ as \cite{yurovsky2015a}
\begin{multline}
\sum_{r,r'}\langle\tilde{\Psi}_{r'\{\mathbf{p}'\}S'}^{(S')}|\hat{V}_{a}|\tilde{\Psi}_{r\{\mathbf{p}\}S}^{(S)}\rangle\langle\tilde{\Psi}_{r'\{\mathbf{p}'\}S'}^{(S')}|\hat{V}_{b}|\tilde{\Psi}_{r\{\mathbf{p}\}S}^{(S)}\rangle^{*}\\
=Y^{(S,2)}[\hat{V}_{a},\hat{V}_{b}]\frac{2f_{S'}}{N(N-1)}\sum_{j<j'}|\langle\mathbf{p}'_{j}\mathbf{p}'_{j'}|\delta|\mathbf{p}_{j}\mathbf{p}_{j'}\rangle|^{2}\\
\times\prod_{j'\neq j''\neq j}\delta_{\mathbf{p}'_{j''}\mathbf{p}_{j''}}\label{eq:SumVaVbnpn}
\end{multline}
Here the universal factors $Y^{(S,2)}[\hat{V}_{a},\hat{V}_{b}]$ are
independent of the occupied spatial modes, while the sum of matrix
elements is independent of the total spin and its projection. The
sum rule \eqref{eq:SumVaVbnpn} is valid if the sets $\{\mathbf{p}\}$
and $\{\mathbf{p}'\}$ are different by two momenta. Similar sum for
$\{\mathbf{p}'\}=\{\mathbf{p}\}$ is proportional to deviations of
the matrix elements $\langle\mathbf{p}'\mathbf{p}|\delta|\mathbf{p}\mathbf{p}'\rangle$
from their average values \cite{yurovsky2015a} and vanish in the
present case since
\[
\langle\mathbf{p}'_{j}\mathbf{p}'_{j'}|\delta|\mathbf{p}_{j}\mathbf{p}_{j'}\rangle=L^{-D}\delta_{\mathbf{p}'_{j}+\mathbf{p}'_{j'}\mathbf{p}_{j}+\mathbf{p}_{j'}}
\]
and the matrix elements $\langle\mathbf{p}'\mathbf{p}|\delta|\mathbf{p}\mathbf{p}'\rangle=L^{-D}$
are constant. The sets $\{\mathbf{p}\}$ and $\{\mathbf{p}'\}$ different
by single momentum are not coupled due to the momentum conservation.

The form \eqref{eq:SumVaVbnpn} of the sum implies that the unchanged
momenta are in the same positions in the sets $\{\mathbf{p}\}$ and
$\{\mathbf{p}'\}$. However, arbitrary permutations $\pr P$ and $\pr P'$
of the momentum sets do not change the sum, 
\begin{multline}
\sum_{r,r'}\langle\tilde{\Psi}_{r'\pr P'\{\mathbf{p}'\}S'}^{(S')}|\hat{V}_{a}|\tilde{\Psi}_{r\pr P\{\mathbf{p}\}S}^{(S)}\rangle\langle\tilde{\Psi}_{r'\pr P'\{\mathbf{p}'\}S'}^{(S')}|\hat{V}_{b}|\tilde{\Psi}_{r\pr P\{\mathbf{p}\}S}^{(S)}\rangle^{*}\\
=\sum_{r,r'}\langle\tilde{\Psi}_{r'\{\mathbf{p}'\}S'}^{(S')}|\hat{V}_{a}|\tilde{\Psi}_{r\{\mathbf{p}\}S}^{(S)}\rangle\langle\tilde{\Psi}_{r'\{\mathbf{p}'\}S'}^{(S')}|\hat{V}_{b}|\tilde{\Psi}_{r\{\mathbf{p}\}S}^{(S)}\rangle^{*}.\label{eq:PermSumME}
\end{multline}
This equality is provided by the transformation of the non-interacting
particle wavefunctions on permutation of the quantum numbers (see
\cite{kaplan})
\[
\tilde{\Psi}_{r\pr P\{\mathbf{p}\}S}^{(S)}=\sum_{r'}D_{r'r}^{[\lambda]}(\pr P)\tilde{\Psi}_{r'\{\mathbf{p}\}S}^{(S)}
\]
and the identity \eqref{eq:ProdInv}. Then the sum rule \eqref{eq:SumVaVbnpn}
can be rewritten as
\begin{multline}
\sum_{r,r'}\langle\tilde{\Psi}_{r'\{\mathbf{p}'\}S'}^{(S')}|\hat{V}_{a}|\tilde{\Psi}_{r\{\mathbf{p}\}S}^{(S)}\rangle\langle\tilde{\Psi}_{r'\{\mathbf{p}'\}S'}^{(S')}|\hat{V}_{b}|\tilde{\Psi}_{r\{\mathbf{p}\}S}^{(S)}\rangle^{*}\\
=Y^{(S,2)}[\hat{V}_{a},\hat{V}_{b}]\frac{f_{S'}}{N(N-1)}\sum_{\pr P}\sum_{j<j'}|\langle\mathbf{p}'_{\pr Pj}\mathbf{p}'_{\pr Pj'}|\delta|\mathbf{p}_{j}\mathbf{p}_{j'}\rangle|^{2}\\
\times\prod_{j'\neq j''\neq j}\delta_{\mathbf{p}'_{\pr Pj''}\mathbf{p}_{j''}}.\label{eq:SumArbSetp}
\end{multline}
The factor $2$ in \prettyref{eq:SumVaVbnpn} is absent here since
the Kronecker symbols are satisfied by two permutations, $\pr P$
and $\pr P\pr P_{jj'}$, when the momenta $\mathbf{p}_{j}$ and $\mathbf{p}_{j'}$
are changed. 

In the matrix element \eqref{eq:MatElPsi2} between the wavefunctions
of interacting particles, we can replace the sum over $\{\mathbf{p}\}$
in the simplex $\mathbf{p}_{1}<\mathbf{p}_{2}<\cdots<\mathbf{p}_{N}$
by average of sums over all simplices, $\mathbf{p}_{\pr P1}<\mathbf{p}_{\pr P2}<\cdots<\mathbf{p}_{\pr PN}$,
\[
{\sum_{\{\mathbf{p}\}}}'F(\{\mathbf{p}\})=\frac{1}{N!}\sum_{\pr P}{\sum_{\pr P\{\mathbf{p}\}}}'F(\pr P\{\mathbf{p}\}).
\]
Since the sums of the matrix elements \eqref{eq:SumArbSetp} and sums
of squared momenta $\{\mathbf{p}\}^{2}$ are invariant over momentum
permutations, \prettyref{eq:MatElPsi2} takes the form
\begin{multline}
\left\langle |\langle\Psi_{n'S'_{z}}^{(S')}|\hat{V}_{\mathrm{spin}}|\Psi_{nS_{z}}^{(S)}\rangle|^{2}\right\rangle _{\mathrm{EE}}=\left(\frac{\mathcal{N}_{n}^{(S)}\mathcal{N}_{n'}^{(S')}}{N!}\right)^{2}\\
\times{\sum_{\{\mathbf{p}\}}}''{\sum_{\{\mathbf{p'}\}}}''\tilde{\delta}(\{\mathbf{p}\}^{2}-2mE_{n}^{(S)})\tilde{\delta}(\{\mathbf{p'}\}^{2}-2mE_{n'}^{(S')})\\
\times\sum_{r,r'}|\langle\tilde{\Psi}_{r'\{\mathbf{p}'\}S'_{z}}^{(S')}|\hat{V}_{\mathrm{spin}}|\tilde{\Psi}_{r\{\mathbf{p}\}S_{z}}^{(S)}\rangle|^{2}.\label{eq:MatElPsi2f}
\end{multline}
where $\sum_{\{\mathbf{p}\}}''$ denotes summation over the sets $\{\mathbf{p}\}$
which do not contain equal momenta.

The Weisskopf-Wigner decay rates \eqref{eq:GammaWW} for transitions
from the $S$-multiplet to the $S'$ one, $\Gamma_{S_{z}}^{(S,S')}$,
calculated with the matrix elements \eqref{eq:MatElPsi2f} of the
operator \eqref{eq:VspinSph}, relations \eqref{eq:WigEcc}, and sum
rules \eqref{eq:SumArbSetp} attain the form\begin{subequations}\label{eq:GammaXY}
\begin{multline}
\Gamma_{S_{z}}^{(S,S-1)}=\biggl[\frac{2}{3}\left(\alpha_{+}X_{S_{z}0}^{(S,S-1,2)}\right)^{2}Y^{(S,2)}[\hat{V}_{-1}^{(2)},\hat{V}_{-1}^{(2)}]\\
+2\sqrt{\frac{2}{3}}\alpha_{+}\alpha_{-}X_{S_{z}0}^{(S,S-1,2)}X_{S_{z}0}^{(S,S-1,1)}Y^{(S,2)}[\hat{V}_{-1}^{(2)},\hat{V}_{-1}]\\
+\left(\alpha_{-}X_{S_{z}0}^{(S,S-1,1)}\right)^{2}Y^{(S,2)}[\hat{V}_{-1},\hat{V}_{-1}]\biggr]\frac{f_{S-1}}{Nf_{S}}\Gamma_{Dd}
\end{multline}
\begin{multline}
\Gamma_{S_{z}}^{(S,S+1)}=\biggl[\frac{2}{3}\left(\alpha_{+}X_{S_{z}0}^{(S+1,S,2)}\right)^{2}Y^{(S+1,2)}[\hat{V}_{-1}^{(2)},\hat{V}_{-1}^{(2)}]\\
+2\sqrt{\frac{2}{3}}\alpha_{+}\alpha_{-}X_{S_{z}0}^{(S+1,S,2)}X_{S_{z}0}^{(S+1,S,1)}Y^{(S+1,2)}[\hat{V}_{-1}^{(2)},\hat{V}_{-1}]\\
+\left(\alpha_{-}X_{S_{z}0}^{(S+1,S,1)}\right)^{2}Y^{(S+1,2)}[\hat{V}_{-1},\hat{V}_{-1}]\biggr]\frac{1}{N}\Gamma_{Dd}
\end{multline}
\begin{equation}
\Gamma_{S_{z}}^{(S,S-2)}=\frac{2}{3}\left(\alpha_{+}X_{S_{z}0}^{(S,S-2,2)}\right)^{2}Y^{(S,2)}[\hat{V}_{-2}^{(2)},\hat{V}_{-2}^{(2)}]\frac{f_{S-2}}{Nf_{S}}\Gamma_{Dd}
\end{equation}
\begin{equation}
\Gamma_{S_{z}}^{(S,S+2)}=\frac{2}{3}\left(\alpha_{+}X_{S_{z}0}^{(S+2,S,2)}\right)^{2}Y^{(S+2,2)}[\hat{V}_{-2}^{(2)},\hat{V}_{-2}^{(2)}]\frac{1}{N}\Gamma_{Dd}
\end{equation}
\end{subequations} Here the factor $\Gamma_{Dd}$ is independent
of $S$, $S'$, and $S_{z}$, as will be explicitly seen below. It
is obtained from Eqs. \eqref{eq:GammaWW}, \eqref{eq:MatElPsi2f},
and \eqref{eq:SumArbSetp} as 
\begin{multline*}
\Gamma_{Dd}=\frac{\pi f_{S}Ng_{Dd}^{2}}{2\hbar L^{2D}}\left(\frac{\mathcal{N}_{n}^{(S)}\mathcal{N}_{n'}^{(S')}}{N!}\right)^{2}\frac{dn^{(S')}(E_{n'}^{(S')})}{dE}|_{E_{n'}^{(S')}=E_{n}^{(S)}}\\
\times{\sum_{\{\mathbf{p}\}\neq\{\mathbf{p'}\}}}''\tilde{\delta}(\{\mathbf{p}\}^{2}-2mE_{n}^{(S)})\tilde{\delta}(\{\mathbf{p'}\}^{2}-2mE_{n'}^{(S')})\\
\times\frac{1}{N(N-1)}\sum_{j<j'}\sum_{\pr P}\delta_{\mathbf{p}'_{\pr Pj}+\mathbf{p}'_{\pr Pj'}\mathbf{p}_{j}+\mathbf{p}_{j'}}\prod_{j'\neq j''\neq j}\delta_{\mathbf{p}'_{\pr Pj''}\mathbf{p}_{j''}}.
\end{multline*}
The Kronecker symbols here fix values of all $\mathbf{p'}$, except
for $\mathbf{p}'_{j}-\mathbf{p}'_{j'}$. The relation \eqref{eq:DenStatesN}
between the density of states and the normalization factor and replacement
summation by interaction lead to
\begin{multline*}
\Gamma_{Dd}=\frac{\pi mg_{Dd}^{2}}{2\hbar L^{2D}N!}\left(\frac{L}{2\pi\hbar}\right)^{D(N+1)}(\mathcal{N}_{n}^{(S)})^{2}Nf_{S}\\
\times\int d^{DN}p\delta(\{\mathbf{p}\}^{2}-2mE_{n}^{(S)})\int d^{D}p'\delta\left(2{p'}^{2}-\frac{(p_{1}-p_{2})^{2}}{2}\right).
\end{multline*}
Calculating the integrals over $\mathbf{p}_{3},\ldots,\mathbf{p}_{N}$
and $\mathbf{p}_{1}+\mathbf{p}_{2}$ with \prettyref{eq:IntDelta}
and using \prettyref{eq:NormFact} for the normalization factor, we
get 
\begin{multline*}
\Gamma_{Dd}=\frac{N\Gamma(DN/2)mg_{Dd}^{2}}{2^{D+2}\pi^{3D/2-1}\Gamma(D(N-1)/2)\hbar^{D+1}L^{D}(mE_{n}^{(S)})^{DN/2-1}}\\
\times\int d^{D}p(mE_{n}^{(S)}-p^{2})^{D(N-1)/2-1}\int d^{D}p'\delta\left({p'}^{2}-p^{2}\right).
\end{multline*}
Since $\int d^{2}p'\delta\left({p'}^{2}-p^{2}\right)=\pi$ and $\int d^{3}p'\delta\left({p'}^{2}-p^{2}\right)=2\pi p$,
integration over $p$ leads to \prettyref{eq:GammaDd}. 

Substitution the factors $X_{S_{z}0}^{(S,S',q)}$ and $Y^{(S,2)}[\hat{V}_{a},\hat{V}_{b}]$
from \cite{yurovsky2015,yurovsky2015a} into Eqs. \eqref{eq:GammaXY}
leads to \refeq{GammaSSz}.
\begin{acknowledgments}
The author gratefully acknowledges useful conversations with N. Davidson
and I. G. Kaplan. 
\end{acknowledgments}

\end{document}